\PassOptionsToPackage{pdfpagelabels=false}{hyperref}
\documentclass[useAMS,usenatbib,letters]{mnras}
\def \eg {e.g.}

\def \ie {i.e.}

\def \lcdm {{\hbox{$\Lambda$CDM}}}
\def \omegam {{\hbox{$\Omega_m$}}}
\def \omegal {{\hbox{$\Omega_\Lambda$}}}
\def \hzero {{\hbox{$H_0$}}}

\def \arcsec {\hbox{$^{\prime\prime}$}}
\def \deg {\hbox{$^\circ$}}

\def \msun {\hbox{${\rm M_\odot}$}}

\def \mfive {\hbox{$M_{500}$}}

\newcommand{\kmsmpc }{\mbox{km s$^{-1}$ Mpc$^{-1}$}}

\newcommand{\jy }{\mbox{Jy}}
\newcommand{\mjyb }{\mbox{mJy beam$^{-1}$}}
\newcommand{\mujyb }{\mbox{$\mu$Jy beam$^{-1}$}}

\newcommand{\uv }{\textit{uv}}

\newcommand{\prefactor }{\textsc{prefactor}}

\newcommand{\wsclean }{\textsc{WSClean}}

\newcommand{\spam }{\textsc{spam}}

\newcommand{\killms }{\textsc{killMS}}
\newcommand{\ddfacet }{\textsc{DDFacet}}

\newcommand{\chandra }{{\em Chandra}}

\newcommand{\gmrt }{GMRT}

\newcommand{\ugmrt }{uGMRT}
\newcommand{\ugmrtE }{upgraded Giant Metrewave Radio Telescope}

\newcommand{\jvla }{JVLA}
\newcommand{\jvlaE }{Jansky Very Large Array}
\newcommand{\lofar }{LOFAR}
\newcommand{\lofarE }{LOw Frequency ARray}

\newcommand{\lotss }{LoTSS}
\newcommand{\lotssE }{LOFAR Two-meter Sky Survey}

\newcommand{\sdss }{SDSS}

\usepackage{xcolor}
\usepackage{graphicx}
\usepackage{subfig}
\usepackage{amssymb}
\usepackage{multirow,bigdelim}
\usepackage[export]{adjustbox}
\defcitealias{botteon18a1758}{B18}
\defcitealias{govoni19}{G19}
\title[A radio bridge in Abell 1758]{A giant radio bridge connecting two clusters in Abell 1758}

\author[Botteon et al.]{A.~Botteon$^{1}$\thanks{E-mail: botteon@strw.leidenuniv.nl}, R.~J.~van Weeren$^{1}$, G.~Brunetti$^{2}$, F.~de Gasperin$^{3}$, H.~T.~Intema$^{1,4}$, \newauthor
E.~Osinga$^{1}$, G.~Di Gennaro$^{1}$, T.~W.~Shimwell$^{5,1}$, A.~Bonafede$^{6,2,3}$, M.~Br\"{u}ggen$^3$, \newauthor
R.~Cassano$^{2}$, V.~Cuciti$^{3}$, D.~Dallacasa$^{6,2}$, F.~Gastaldello$^{7}$, S.~Mandal$^{1}$, \newauthor
M.~Rossetti$^{7}$ and H.~J.~A.~R\"{o}ttgering$^{1}$ \\
$^{1}$Leiden Observatory, Leiden University, PO Box 9513, 2300 RA Leiden, The Netherlands \\
$^{2}$INAF - IRA, via P.~Gobetti 101, I-40129 Bologna, Italy \\
$^{3}$Hamburger Sternwarte, Universit\"{a}t Hamburg, Gojenbergsweg 112, D-21029 Hamburg, Germany \\
$^{4}$International Centre for Radio Astronomy Research - Curtin University, GPO Box U1987, Perth, WA 6845, Australia \\
$^{5}$ASTRON, the Netherlands Institute for Radio Astronomy, Postbus 2, NL-7990 AA Dwingeloo, The Netherlands \\
$^{6}$Dipartimento di Fisica e Astronomia, Universit\`{a} di Bologna, via P.~Gobetti 93/2, I-40129 Bologna, Italy \\
$^{7}$INAF - IASF Milano, Via A. Corti 12, I-20133, Milano, Italy \\
}
\date{\today}

\date{Accepted XXX. Received YYY; in original form ZZZ}

\pubyear{2020}

\begin{document}
\label{firstpage}
\pagerange{\pageref{firstpage}--\pageref{lastpage}}
\maketitle

\begin{abstract}
Collisions between galaxy clusters dissipate enormous amounts of energy in the intra-cluster medium (ICM) through turbulence and shocks. In the process, Mpc-scale diffuse synchrotron emission in form of radio halos and relics can form. However, little is known about the very early phase of the collision. We used deep radio observations from 53 MHz to 1.5 GHz to study the pre-merging galaxy clusters A1758N and A1758S that are $\sim2$ Mpc apart. We confirm the presence of a giant bridge of radio emission connecting the two systems that was reported only tentatively in our earlier work. This is the second large-scale radio bridge observed to date in a cluster pair. The bridge is clearly visible in the \lofar\ image at 144~MHz and tentatively detected at 53~MHz. Its mean radio emissivity is more than one order of magnitude lower than that of the radio halos in A1758N and A1758S. Interestingly, the radio and X-ray emissions of the bridge are correlated. Our results indicate that non-thermal phenomena in the ICM can be generated also in the region of compressed gas in-between infalling systems.
\end{abstract}


\section{Introduction}

\begin{figure*}
 \centering
 \includegraphics[width=.45\hsize,trim={0cm 0cm 0cm 0cm},clip]{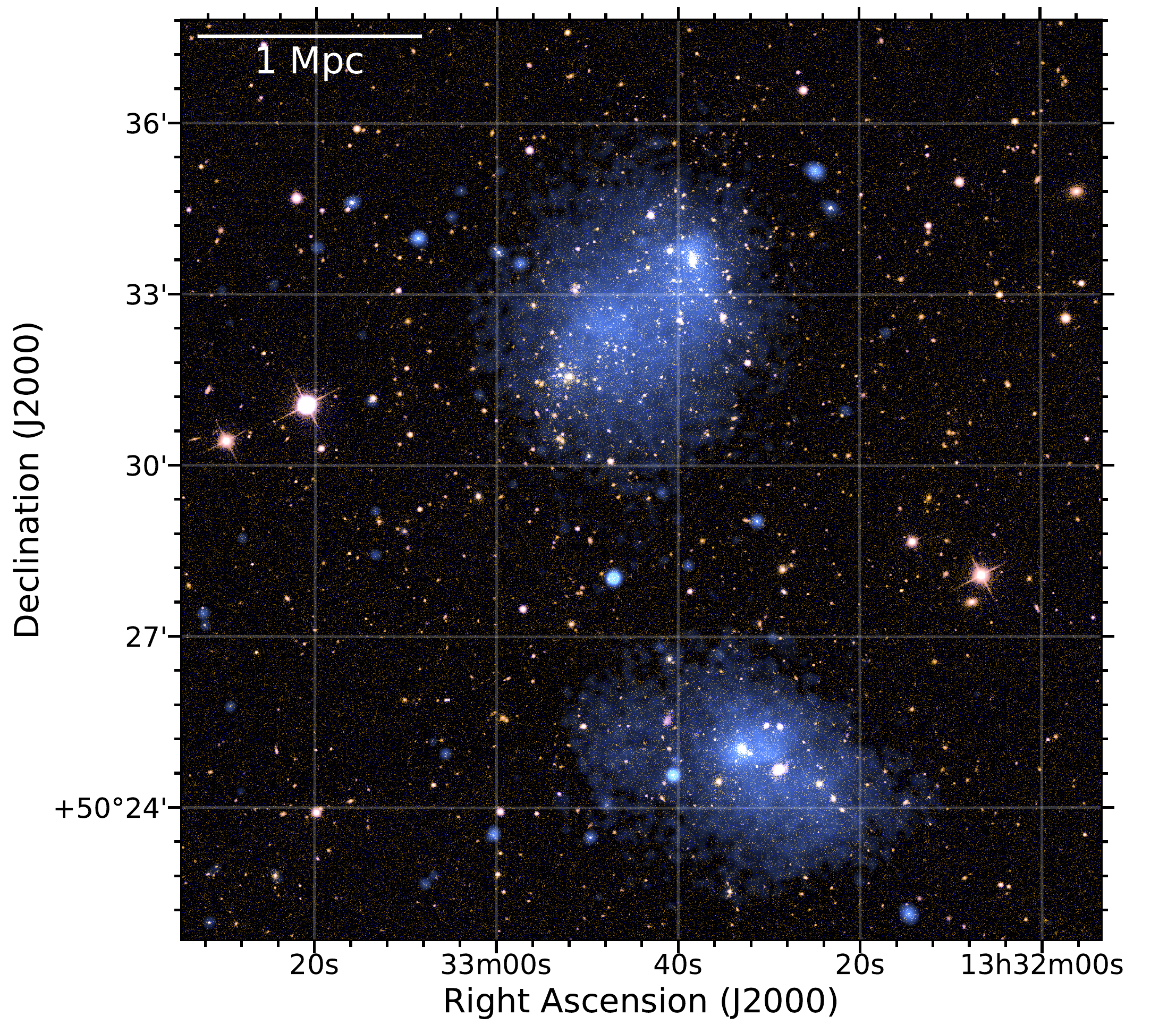}
 \includegraphics[width=.45\hsize,trim={0cm 0cm 0cm 0cm},clip]{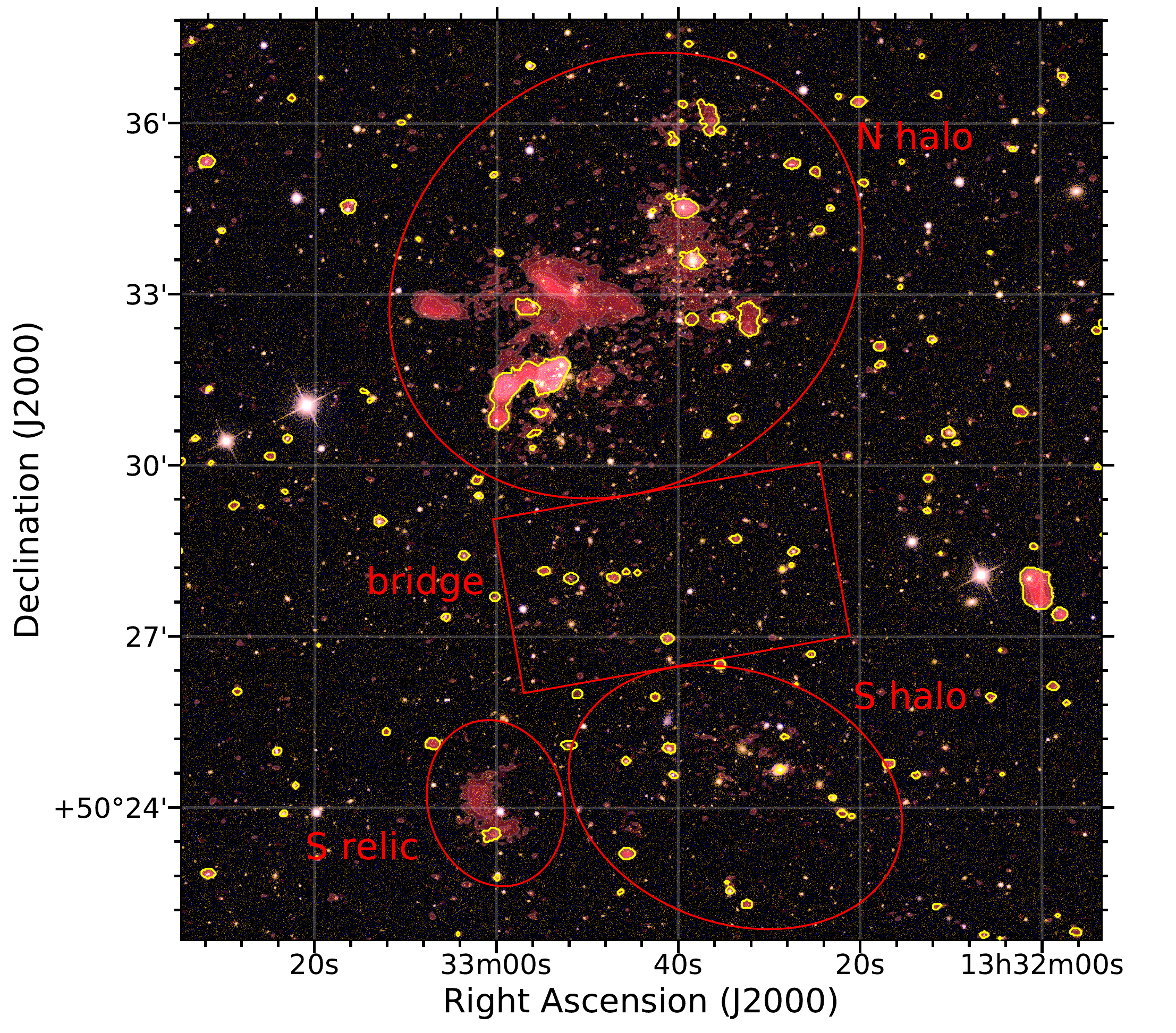}
  \caption{Composite images of A1758 obtained from the superposition of an optical \sdss\ \textit{g,r,i} mosaic with \chandra\ (\textit{blue}) and with a \lofar\ image at 144 MHz with a resolution of $7.6\arcsec \times 5.4\arcsec$ (\textit{red}) and a rms noise of 60 \mujyb. Yellow and red regions indicate the mask used to subtract discrete sources and the regions adopted to measure the flux densities in the low-resolution images, respectively.}
 \label{fig:composite}
\end{figure*}

\begin{table*}
 \scriptsize
 \centering
 \caption{Radio observations used in this work. The offsets of A1758 from \lotss\ pointing centers are: $^a$1.10\deg, $^b$1.70\deg, $^c$2.07\deg, and $^d$2.79\deg.}\label{tab:observations}
 \begin{tabular}{ccrccccccc} 
  \hline
  Telescope & & & & Project & Observation date & Duration & Frequency band & Configuration \\
  \hline
  \multirow{6}*{\lofar} & \ldelim[{6}{*} \hspace{-0.5cm} & & & LC12\_006 & 2020-02-25 & 8~hr & $39-78$ MHz & LBA Outer \\
  & & & & LC12\_006 & 2019-09-29 & 8~hr & $120-168$ MHz & HBA Dual Inner \\
  & & \multirow{4}*{\rotatebox{90}{\lotss}}\hspace{-0.5cm} & \ldelim[{4}{*} \hspace{-0.5cm} & P42Hetdex07$^a$ & 2014-06-01 & 8~hr & $120-168$ MHz & HBA Dual Inner \\
  & & & & P5Hetdex41$^b$ &  2014-07-28 & 8~hr & $120-168$ MHz & HBA Dual Inner \\
  & & & & P38Hetdex07$^c$ & 2014-06-01 & 8~hr & $120-168$ MHz & HBA Dual Inner \\
  & & & & P206$+$52$^d$ & 2015-04-28 & 8~hr & $120-168$ MHz & HBA Dual Inner \\

  \multirow{1}*{\ugmrt} & \ldelim[{1}{*} \hspace{-0.5cm} & &  & 37\_030 & 2019-02-17/18 & 20~hr & $300-500$ MHz & Band 3 \\
  
  \multirow{2}*{\jvla} & \ldelim[{2}{*} \hspace{-0.5cm} & &  & 18A-172 & 2018-09-17/20 & 1.1~hr & $1-2$ GHz & L-band D-array \\
  & & & & 18A-172 & 2019-01-18/19 & 1.1~hr & $1-2$ GHz & L-band C-array \\
  
  \hline
 \end{tabular}
\end{table*}

In the past two decades, the presence of diffuse and extended synchrotron sources with steep spectra ($\alpha>1$, with $S_\nu \propto \nu^{-\alpha}$) in merging galaxy clusters has been confirmed by numerous radio observations \citep[\eg][for a recent review]{vanweeren19rev}. Radio halos (in cluster centers) and radio relics (in cluster outskirts) are among the largest ($\sim$ Mpc-scale) and most common sources associated with the intra-cluster medium (ICM). Their origin is likely related to the process of cluster formation, where part of the energy dissipated into the ICM by turbulence and shocks can be channeled into non-thermal components, namely relativistic particles and magnetic fields \citep[\eg][for a review]{brunetti14rev}. \\
\indent
Highly sensitive observations at low frequencies with \lofarE\ (\lofar) are providing many new insights into the study of non-thermal phenomena in galaxy clusters. Recently, \citet{govoni19} observed a $\sim3$ Mpc radio bridge connecting the pre-merging system Abell 399-401 ($z=0.07$), showing that detectable non-thermal emission can be generated on scales larger than that of clusters. \\
\indent
Abell 1758 (hereafter A1758, see Fig.~\ref{fig:composite}) is a system located at $z=0.279$ composed of two massive galaxy clusters separated by a projected distance of $\sim2$ Mpc: A1758N (in the north, the most massive one) and A1758S (in the south). X-ray observations suggest that the two clusters are gravitationally bound but have not interacted yet, that is, they are in a pre-merging phase \citep{david04, botteon18a1758, schellenberger19}. In addition, complex cluster dynamics and multiple sub-substructures are observed both in A1758N and A1758S, indicating that each of the two clusters is undergoing its own merger \citep[\eg][]{monteirooliveira17a1758}. In \citet{botteon18a1758}, we used 144~MHz \lofar\ observations to study the well-known radio halo in A1758N and also discovered a new radio halo and a candidate radio relic in A1758S. More importantly, at low resolution we found a hint ($2\sigma$) of a bridge of radio emission connecting the two clusters which required a further study with more sensitive observations. \\
\indent
In this Letter, we report the results of an extensive campaign of deep, multi-frequency, radio observations of the radio bridge connecting the galaxy clusters in A1758. Here, we adopt a \lcdm\ cosmology with $\omegal = 0.7$, $\omegam = 0.3$ and $\hzero = 70$ \kmsmpc, in which 1 arcsec corresponds to 4.233 kpc at the cluster redshift ($z=0.279$) and the luminosity distance is $D_L = 1428$ Mpc.

\begin{figure*}
 \centering
 \includegraphics[width=.49\hsize,trim={0cm 0cm 0cm 0cm},clip]{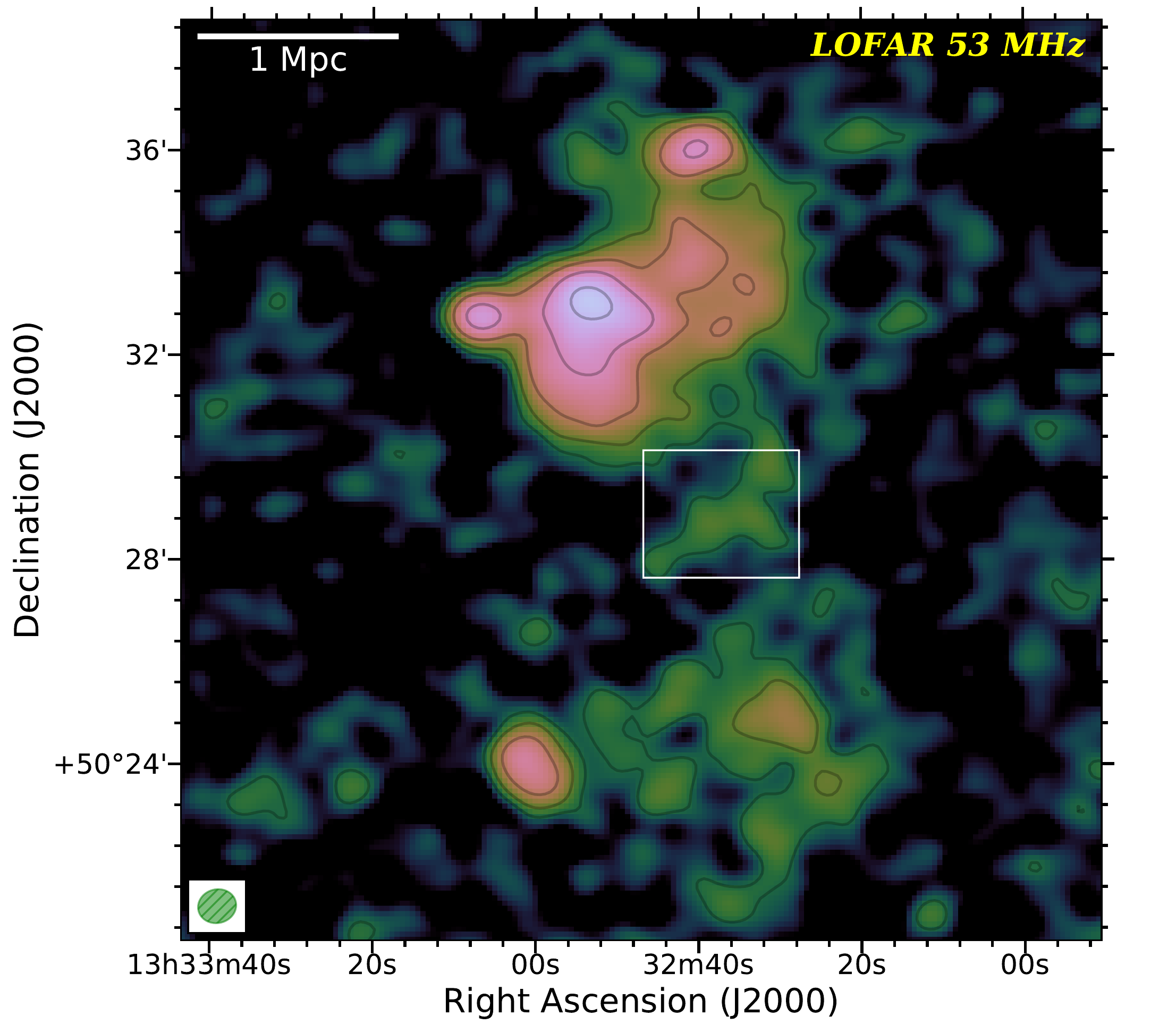}
 \includegraphics[width=.49\hsize,trim={0cm 0cm 0cm 0cm},clip]{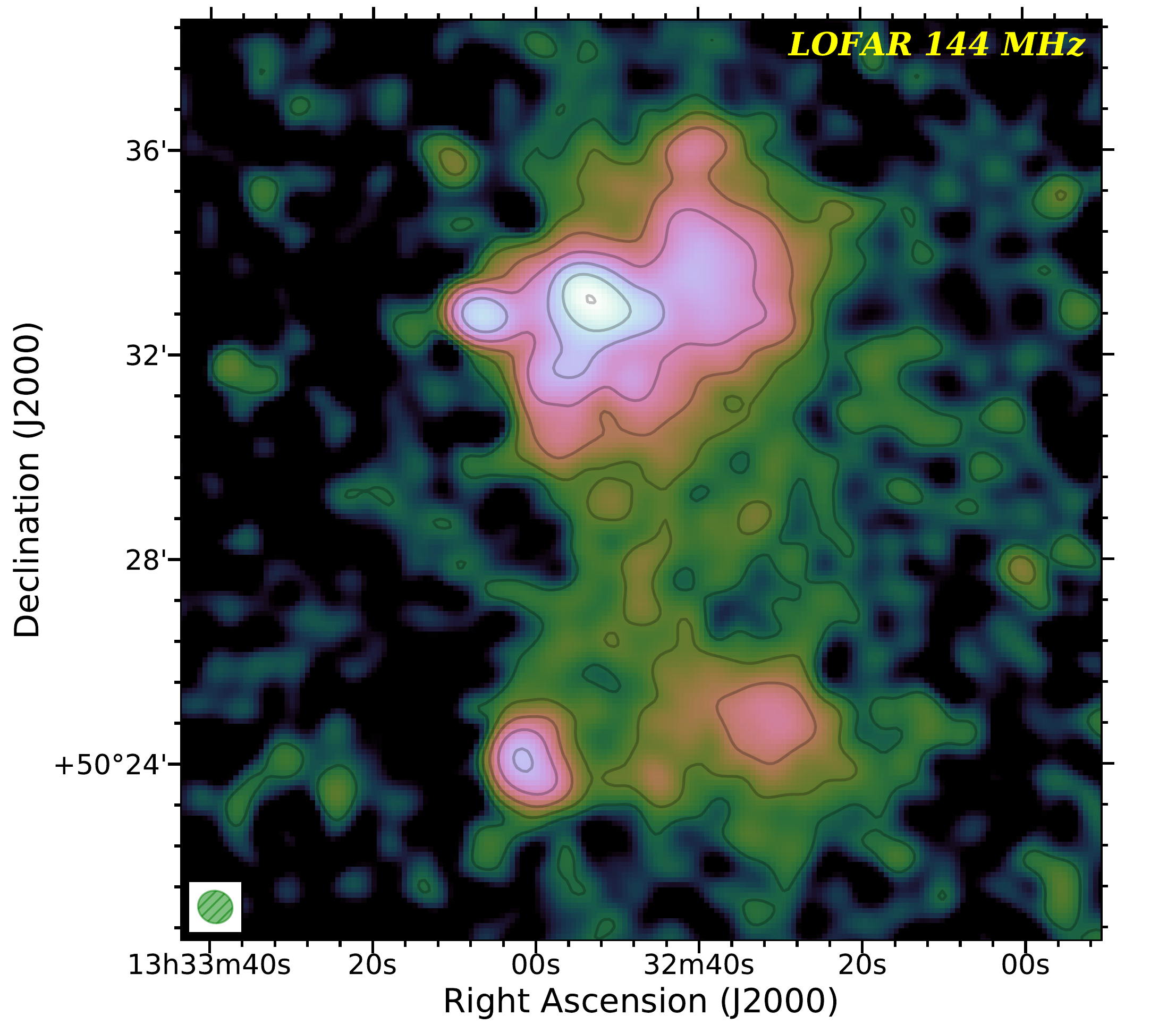}
 \includegraphics[width=.49\hsize,trim={0cm 0cm 0cm 0cm},clip]{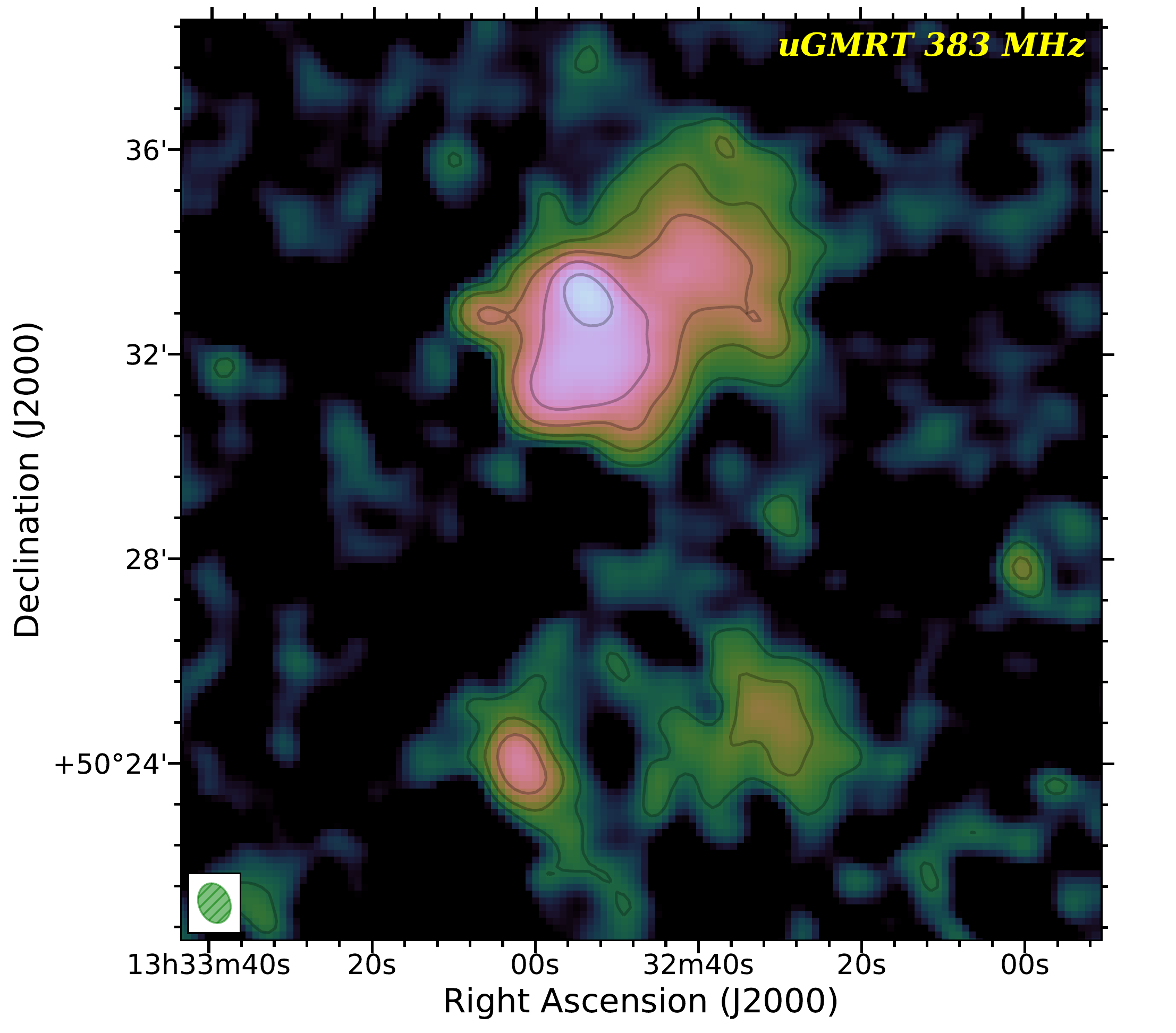}
 \includegraphics[width=.49\hsize,trim={0cm 0cm 0cm 0cm},clip]{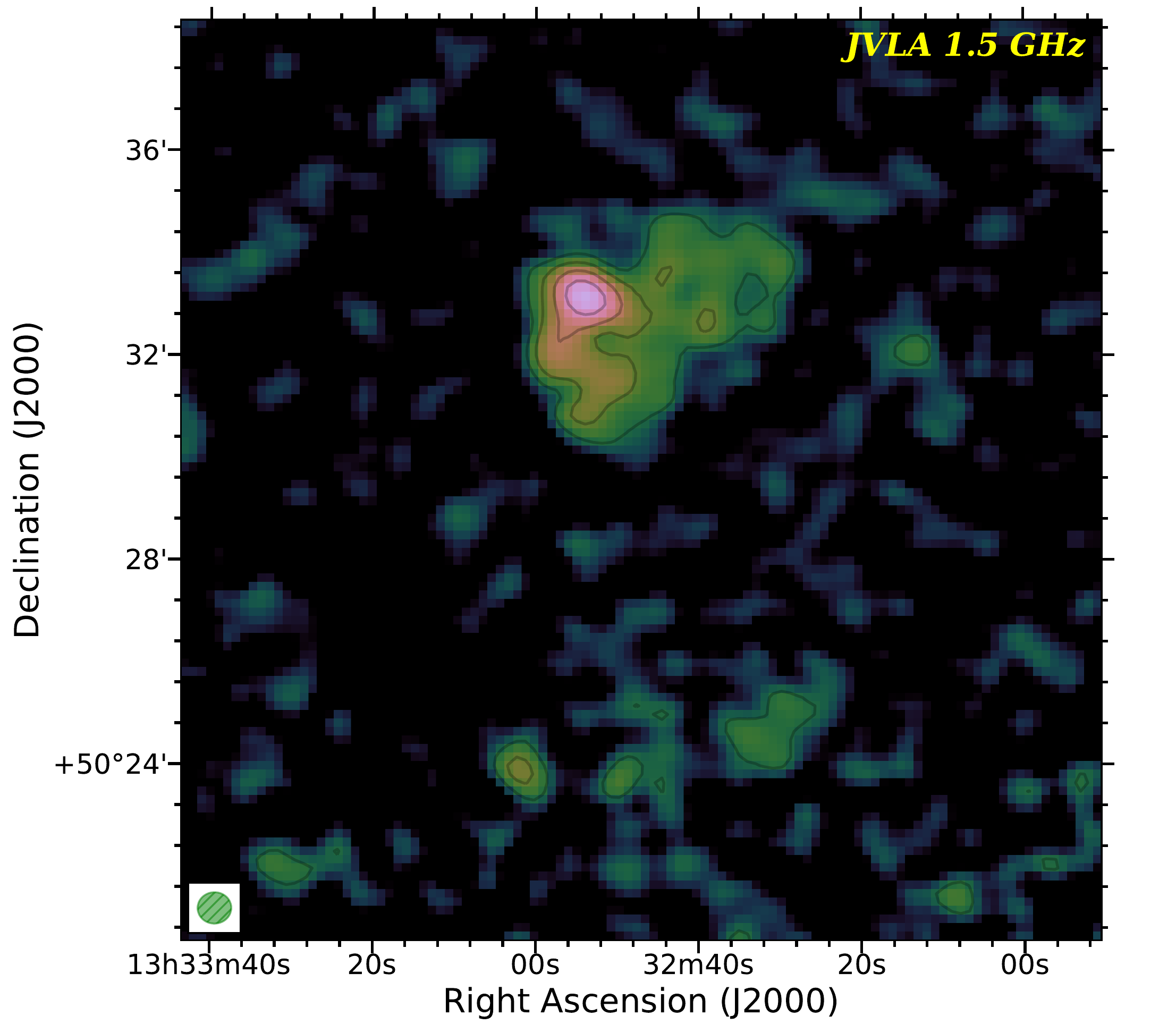}
 \caption{Radio images from 53 MHz to 1.5 GHz of A1758 with discrete sources subtracted. The color scale has a logarithmic stretch from 0.5 to 200$\sigma$. Contours are drawn from $3\sigma$ and are spaced by a factor of 2, where $\sigma_{\rm{53\,MHz}} = 1.6$ \mjyb, $\sigma_{\rm{144\,MHz}} = 160$ \mujyb, $\sigma_{\rm{383\,MHz}} = 170$ \mujyb, $\sigma_{\rm{1.5\,GHz}} = 80$ \mujyb. Images were obtained by applying a Gaussian \uv-taper of 35\arcsec; restoring beams are shown in the bottom left corners. The white box denotes the region where we measured a spectral index of $\alpha = 1.65\pm0.27$ between 53 and 144 MHz.}
 \label{fig:multifreq}
\end{figure*}

\section{Observations and data reduction}

We observed A1758 with the \lofar\ Low/High Band Antenna (LBA/HBA) arrays, the \ugmrtE\ (\gmrt), and the \jvlaE\ (\jvla). Details of our observations are summarized in Tab.~\ref{tab:observations}. The data reduction procedures for each dataset are briefly described below. For all calibrated datasets, the final imaging has been performed using \wsclean\ v2.8 \citep{offringa14} with multi-scale multi-frequency deconvolution \citep{offringa17}.

\subsection{\lofar}

In \citet{botteon18a1758}, we analyzed the pointing closest to A1758 (offset by $\sim1.1\deg$) coming from the \lotssE\ \citep[\lotss;][]{shimwell19}. For this work, we exploit follow-up observations with \lofar\ LBA ($39-78$ MHz) and HBA ($120-168$ MHz) in combination with four \lotss\ pointings that lay within $3\deg$ of A1758. For LBA, the target and calibrator were jointly observed for 8~hr using the multi-beam capability of \lofar, while HBA observations followed the scheme of \lotss, namely 8~hr runs book-ended by 10 min scans on flux density calibrators. Each observation was analyzed individually with the pipelines developed by the \lofar\ Surveys Key Science Project team (\prefactor, \citealt{degasperin19}; \killms, \citealt{tasse14arx, smirnov15}; \ddfacet, \citealt{tasse18}) to correct for direction-independent effects and to perform a first round of direction-dependent calibration of the entire \lofar\ field-of-view before combination. The image quality towards A1758 was improved following the scheme that has been adopted in recent \lofar\ HBA works (\eg, \citealt{botteon19lyra, botteon20a2255}; van Weeren et al.,~in prep.), which consists of the subtraction of the sources outside the target region from the visibility data, phase-shifting to the center of the region, and correcting the \lofar\ station beam towards this direction. Residual artifacts are reduced by means of phase and amplitude self-calibration loops in the small region containing the target. The same procedure was adapted with optimized parameters for the LBA data. We set conservative systematic uncertainities of 15\% and 20\% on 53~MHz (LBA) and 144~MHz (HBA) flux densities, respectively.

\subsection{\ugmrt}

We have observed A1758 for 20~hr in band 3 ($300-500$ MHz) with the \ugmrt. Data were recorded in 2048 frequency channels with integration time of 4~s in full Stokes mode. The dataset was split into six frequency slices with a bandwidth of 33.3 MHz centered from 317 to 482 MHz that were processed independently using the \spam\ pipeline \citep{intema09}. In the final analysis we removed the highest frequency sub-band due to its lower quality and jointly deconvolved the remaining five slices to produce images with a central frequency at 383 MHz. The systematic uncertainity due to residual amplitude errors was set to 15\%.

\subsection{\jvla}

The \jvla\ L-band ($1-2$ GHz) observations consist of two 1.1~hr runs performed with the C and D arrays. Each dataset was reduced following standard procedures, including removal of the radio frequency interference, calibration of antenna delays and positions, bandpass, cross-hand delays, and polarization leakage and angle. To optimize the image quality, we performed various iterations of self-calibration to calculate the calibration solutions. First, three rounds of phase and three rounds of amplitude calibrations are done on the C and D array data separately. Then, another six rounds of self-calibration were performed on the combined C+D array to produce the final dataset with central frequency 1.5~GHz. The absolute flux scale calibration error was set to 5\%.

\section{Results}

\begin{figure}
 \centering
 \includegraphics[width=.9\hsize,trim={0cm 0cm 0cm 0cm},clip]{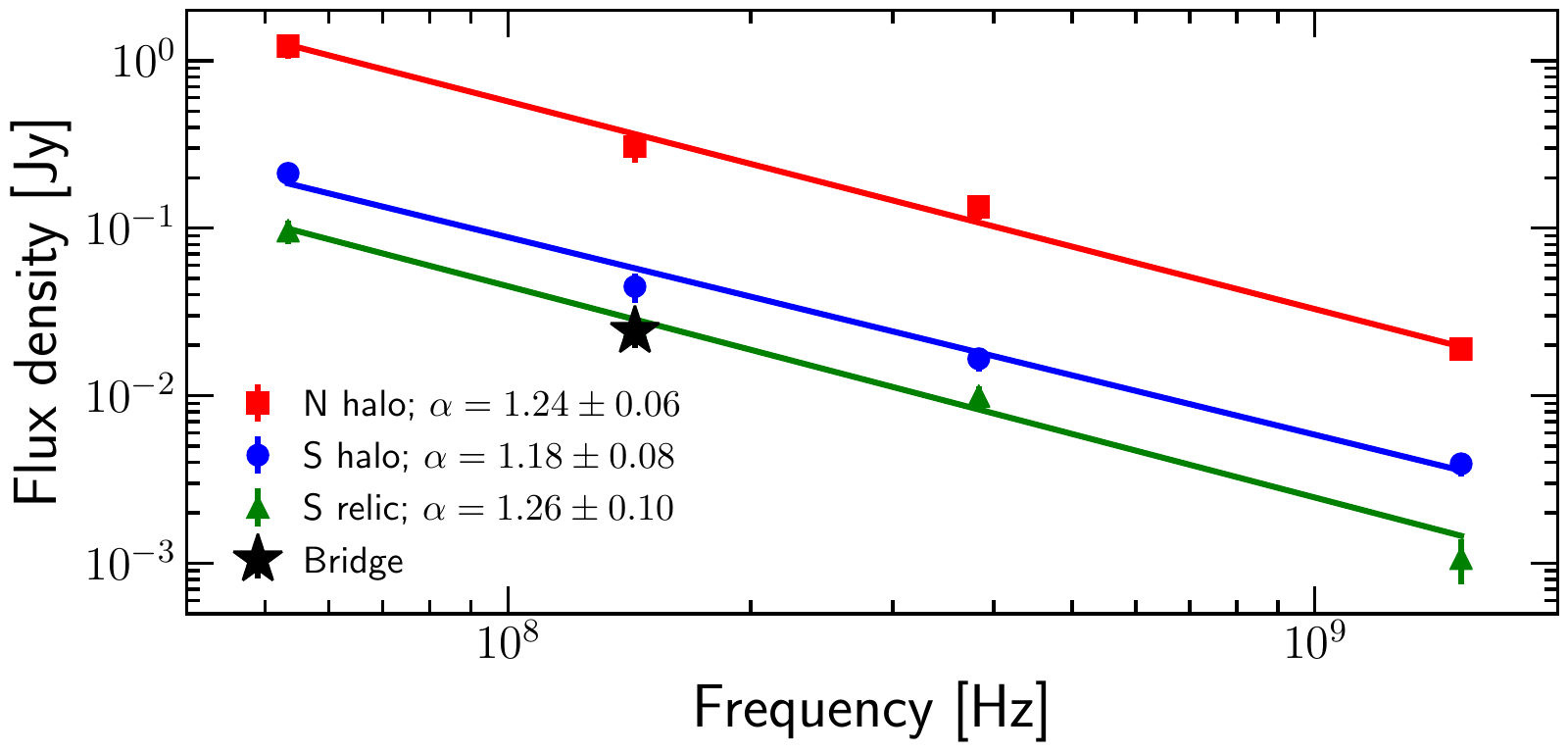}
  \caption{Integrated spectra of the diffuse radio sources.}
 \label{fig:spectra}
\end{figure}

We produced deep images of the A1758 system using the observations listed in Tab.~\ref{tab:observations}. Our \lofar\ HBA image in Fig.~\ref{fig:composite} is a factor of $\sim2$ deeper than that published in \citet{botteon18a1758} as a result of the combination of different observations and recent improvements in the data reduction pipelines. In order to properly study the extended emission from the ICM and determine its flux density, a careful subtraction of the emission from discrete sources embedded in the cluster emission must be performed. For this reason, we followed the approach reported in \citet{botteon18a1758}, in which different models of point sources are subtracted from the visibilities to assess the accuracy of the procedure. Our discrete sources models were created by making 6 images with inner \uv-cuts equally spaced in the range $1.0-3.5$ k$\lambda$ (equivalent to $873-249$ kpc at the redshift of A1758) for the \lofar\ HBA and LBA, \ugmrt\ and \jvla\ datasets deploying custom cleaning masks (\eg\ Fig.~\ref{fig:composite}). In this phase, multiscale cleaning was switched off to minimize the amount of diffuse emission from the ICM picked up by the deconvolution algorithm. The aforementioned models were subtracted individually from the visibilities and six images of the diffuse emission were produced with the same \wsclean\ parameters for each dataset. An example of low-resolution images from 53 MHz to 1.5 GHz with comparable restoring beams are reported in Fig.~\ref{fig:multifreq}. In the following, quoted flux densities at a given frequency represent the median value of the six different source-subtracted images measured in the same regions (the relative standard deviation of the six images is typically $\lesssim 2 \%$). \\
\indent
The presence of diffuse radio sources in A1758N and A1758S is confirmed from 53 MHz to 1.5 GHz (Fig.~\ref{fig:multifreq}). Most remarkably, we clearly observe a bridge of emission connecting the two clusters in the \lofar\ image at 144 MHz, where the emission fills the region between A1758N and A1758S and has an integrated flux density of $24.2\pm4.9$ m\jy. Low-significance patches of emission are observed both at 53 and 383 MHz, the most significant of which is a filamentary structure extending from the halo in A1758N towards A1758S detected above the $3\sigma$ level at 53 MHz. This region is encompassed by the white box shown in Fig.~\ref{fig:multifreq} (top-left panel) where we measure flux densities of $S_{\rm{53\,MHz}} = 60.2\pm10.7$ m\jy\ and $S_{\rm{144\,MHz}} = 11.6\pm2.4$ m\jy, leading to $\alpha=1.65\pm0.27$. By assuming a limit at $3\sigma$ on the 53~MHz flux density, the average spectral index on the entire bridge results $\alpha < 1.84$, while current \ugmrt\ and \jvla\ data provide only a loose constraint of $\alpha>0.4$. \\
\indent
The flux densities and integrated spectra of the other diffuse sources in the system measured from the red regions shown in Fig.~\ref{fig:composite} (which were drawn to roughly follow the $3\sigma$ level contour of the \lofar\ HBA image in Fig.~\ref{fig:multifreq}) are reported in Fig.~\ref{fig:spectra}. Our measurements agree with the results of \citet{botteon18a1758} and \citet{schellenberger19}.

\section{Discussion}

\begin{figure}
 \centering
 \begin{tabular}{cc}
  \vspace{-0.10cm}\hspace{-0.40cm} \multirow{2}{*}{\subfloat{\includegraphics[width=.61\hsize,trim={0.9cm 0.1cm 0.9cm 0.1cm},clip]{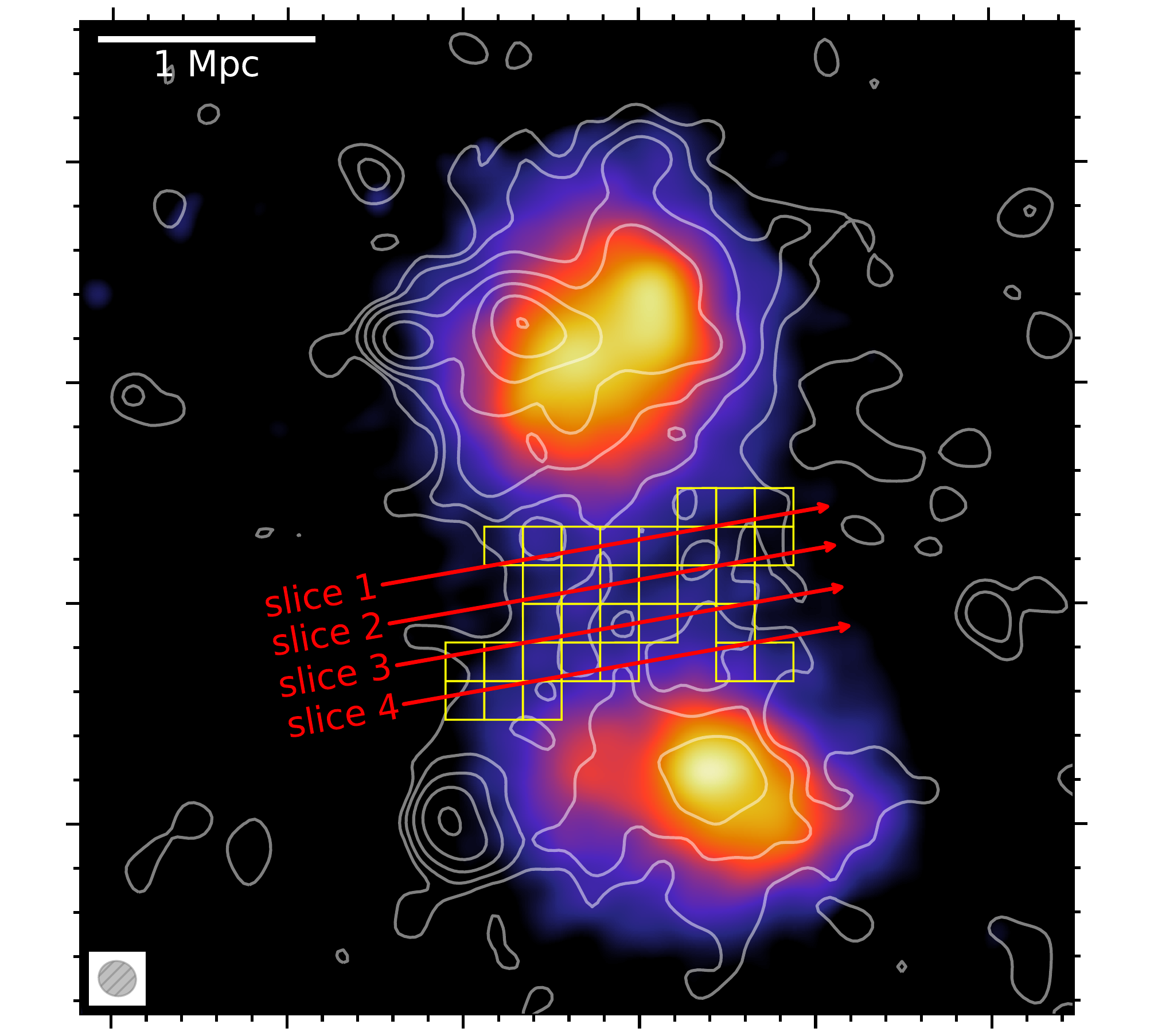}}} & \vspace{-0.25cm} \\
  & \vspace{-0.25cm}\hspace{-0.4cm}\subfloat{\includegraphics[width=.39\hsize]{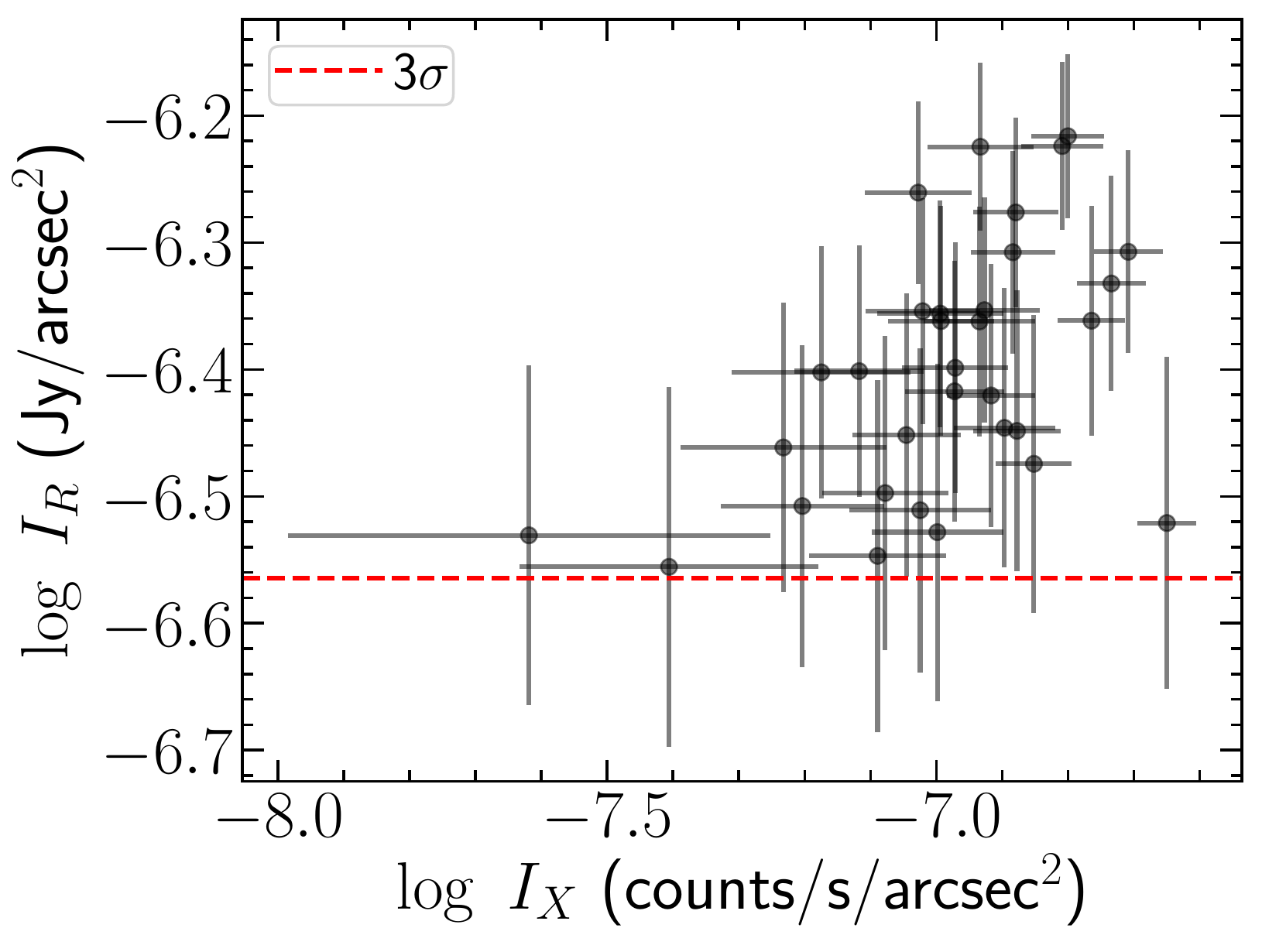}} \\
  & \vspace{-0.1cm}\hspace{-0.4cm}\subfloat{\includegraphics[width=.39\hsize]{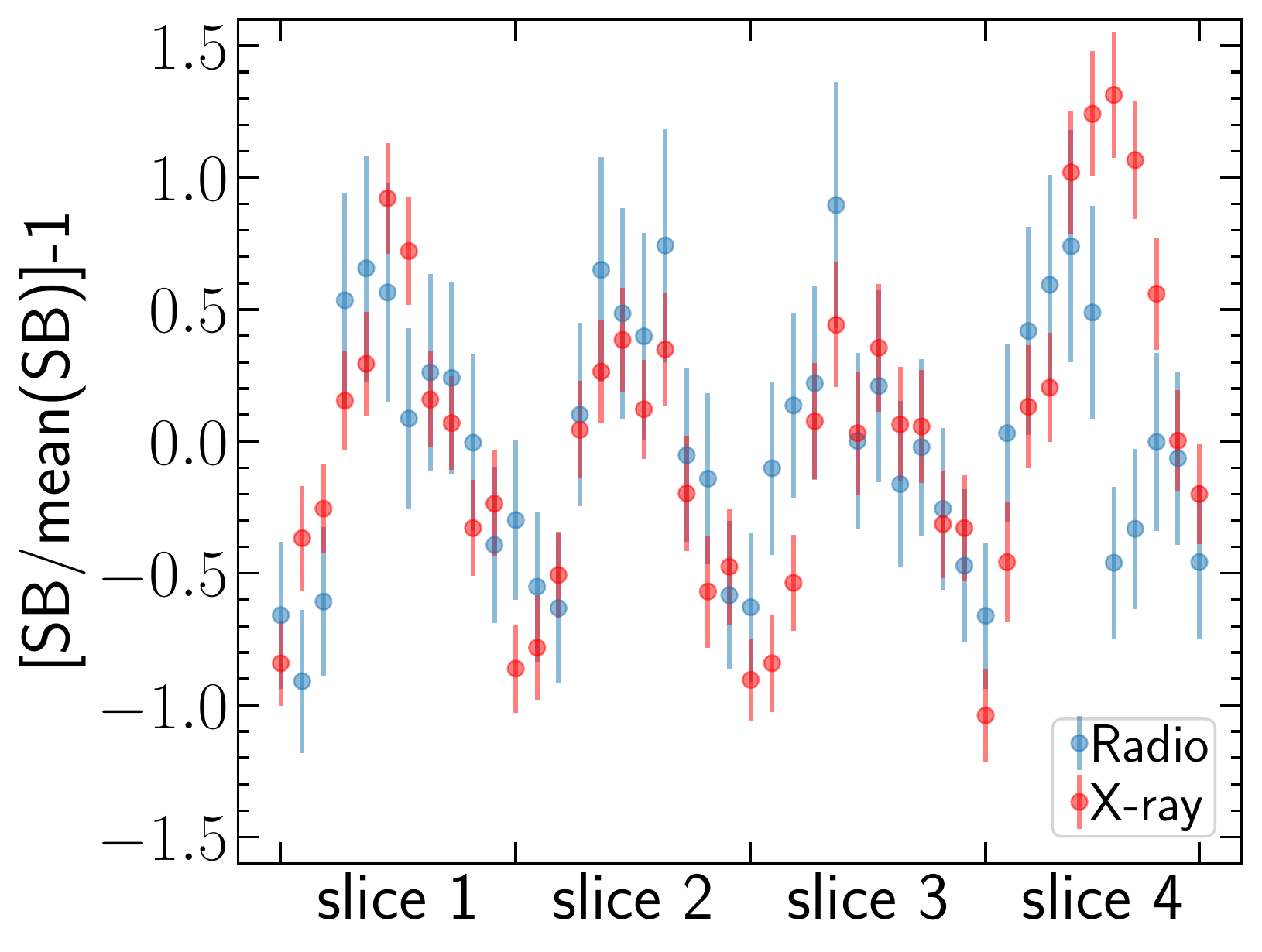}} \\
 \end{tabular}
  \caption{\chandra\ smoothed image in the $0.5-2.0$ keV band overlaid with the low-resolution \lofar\ HBA contours (\textit{left}). The boxes and slices indicate the regions where the X-ray and radio surface brightness have been evaluated and compared (\textit{right}).}
 \label{fig:overlay}
\end{figure}

Radio bridges connecting pre-merging clusters are a recent discovery. So far, the cluster pairs A1758N-A1758S (at $z = 0.279$) and Abell 399-401 \citep[at $z = 0.07$,][]{govoni19} are the only two cases where a bridge of radio emission between two clusters has been observed. The two systems show remarkable similarities. First of all, each of the two main components of the pairs is a massive cluster, with $\mfive \gtrsim 5\times10^{14}$ \msun\ \citep{planck16xxvii}. Second, both are pairs of dynamically disturbed clusters, with all four clusters undergoing mergers and hosting a radio halo \citep{murgia10, botteon18a1758}. Conversely, recent \lofar\ HBA observations failed to detect a radio bridge connecting the two clusters in the Lyra complex \citep{botteon19lyra}, that are less massive and in that case only one of the two is a merging system/hosts a radio halo. This may suggest that radio bridges form from the dissipation of energy in dynamically active regions. \\
\indent
According to \citet{brunetti20}, radio bridges may originate from second-order Fermi acceleration of electrons interacting with turbulent motions triggered by the complex dynamics in the overdense region between pre-merging clusters. The observation of infalling sub-groups onto A1758 \citep{haines18, schellenberger19} would be in agreement with this scenario. One of the key predictions of this model is that the emission in radio brides is volume-filling especially at low frequencies. In this respect, the detection at 144~MHz combined with the X-ray observations provides important information. In Fig.~\ref{fig:overlay} we compare the X-ray and radio surface brightness of the bridge extracted in 31 regions and along 4 transverse slices, finding a remarkably good correlation between the two. We observe fluctuations of similar magnitude and morphology in radio and X-rays, suggesting that thermal and non-thermal emissions are connected and originate from similar volumes. This connection is studied for the first time in a radio bridge and is in line with theoretical expectations. The spectrum of radio bridges also provides important information on their origin. Unfortunately, we were able to provide a spectral constraint only in a small region of the bridge, that may be not representative of the overall spectrum of the emission. Deeper observations are required for this kind of study. Based on the red regions in Fig.~\ref{fig:composite}, we assume oblate spheroid geometries for the halos and a cylindrical volume for the bridge to compute the average radio emissivity of the diffuse sources. The mean radio emissivity of the bridge is $\langle \epsilon\rangle_{\rm{144\,MHz}}  = 4.02 \times 10^{-43}$ erg s$^{-1}$ Hz$^{-1}$ cm$^{-3}$, that is more than one order of magnitude lower than that of the two halos. We note that the emissivity of the bridge in A1758 is also similar (a factor of 2 lower) to that reported for the bridge in Abell 399-401 \citep{govoni19}, marking another similarity between the two pairs. \\
\indent
Regions between clusters give us the possibility to study the most diluted regions of the ICM that are accessible with current instruments \citep{vazza19}. They are dynamically very young, \ie\ their dynamical age is comparable to the eddy turnover time of the turbulent eddies generated in these environments. Under these conditions, the media are characterized by a ratio of thermal to magnetic pressures reasonably $> 100$ (in clusters we believe that $P_{\rm th}/P_B = 10 - 100$), providing unique laboratories to study magnetic field amplification and particle acceleration in new regimes \citep[see][]{brunetti20}.

\section{Conclusions}

We have confirmed the presence of a $\sim2$ Mpc radio bridge connecting the two galaxy clusters in A1758. A standalone detection could be claimed only at 144~MHz, where the radio and X-ray emissions are correlated, suggesting that they share similar emitting volumes. Only hints of radio emission are observed at 53 and 383 MHz, making uncertain the determination of its spectral index; deeper observations are required to provide a robust estimate of its value. \\
\indent
Only two giant intra-cluster radio bridges have been detected to date. These are among the most giant structures observed in the Universe so far, and their origin is likely related to the turbulence (and shocks) generated in the ICM during the initial stage of the merger, which boost both the radio and X-ray emission between the clusters. These detections demonstrate the existence of non-thermal components at large distances from cluster centers with important implications for the models of magnetic field amplification and particle acceleration in the most diluted regions of the ICM. 

\section*{Acknowledgments}

ABot, RJvW, and EO acknowledge support from the VIDI research programme with project number 639.042.729, which is financed by the Netherlands Organisation for Scientific Research (NWO). GDG acknowledges support from the ERC Starting Grant ClusterWeb 804208. ABon acknowledges support from the ERC-Stg DRANOEL n. 714245 and from the MIUR FARE grant ``SMS''. GB, RC, FG, and MR acknowledge support from INAF mainstream project ``Galaxy Clusters Science with LOFAR'' 1.05.01.86.05. VC acknowledges support from the Alexander von Humboldt Foundation. 
LOFAR \citep{vanhaarlem13} is the Low Frequency Array designed and constructed by ASTRON. It has observing, data processing, and data storage facilities in several countries, which are owned by various parties (each with their own funding sources), and that are collectively operated by the ILT foundation under a joint scientific policy. The ILT resources have benefited from the following recent major funding sources: CNRS-INSU, Observatoire de Paris and Universit\'{e} d'Orl\'{e}ans, France; BMBF, MIWF-NRW, MPG, Germany; Science Foundation Ireland (SFI), Department of Business, Enterprise and Innovation (DBEI), Ireland; NWO, The Netherlands; The Science and Technology Facilities Council, UK; Ministry of Science and Higher Education, Poland; The Istituto Nazionale di Astrofisica (INAF), Italy. This research made use of the Dutch national e-infrastructure with support of the SURF Cooperative (e-infra 180169) and the LOFAR e-infra group. The J\"{u}lich LOFAR Long Term Archive and the German LOFAR network are both coordinated and operated by the J\"{u}lich Supercomputing Centre (JSC), and computing resources on the supercomputer JUWELS at JSC were provided by the Gauss Centre for Supercomputing e.V. (grant CHTB00) through the John von Neumann Institute for Computing (NIC). This research made use of the University of Hertfordshire high-performance computing facility and the LOFAR-UK computing facility located at the University of Hertfordshire and supported by STFC [ST/P000096/1], and of the Italian LOFAR IT computing infrastructure supported and operated by INAF, and by the Physics Department of Turin University (under an agreement with Consorzio Interuniversitario per la Fisica Spaziale) at the C3S Supercomputing Centre, Italy. We thank the staff of the GMRT for support. GMRT is run by the National Centre for Radio Astrophysics of the Tata Institute of Fundamental Research. The NRAO is a facility of the National Science Foundation operated under cooperative agreement by Associated Universities, Inc.

\section*{Data availability}

The data underlying this article will be shared on reasonable request to the corresponding author.

\bibliographystyle{mnras}
\bibliography{library.bib}

\begin{thebibliography}{}
\makeatletter
\relax
\def\mn@urlcharsother{\let\do\@makeother \do\$\do\&\do\#\do\^\do\_\do\%\do\~}
\def\mn@doi{\begingroup\mn@urlcharsother \@ifnextchar [ {\mn@doi@}
  {\mn@doi@[]}}
\def\mn@doi@[#1]#2{\def\@tempa{#1}\ifx\@tempa\@empty \href
  {http://dx.doi.org/#2} {doi:#2}\else \href {http://dx.doi.org/#2} {#1}\fi
  \endgroup}
\def\mn@eprint#1#2{\mn@eprint@#1:#2::\@nil}
\def\mn@eprint@arXiv#1{\href {http://arxiv.org/abs/#1} {{\tt arXiv:#1}}}
\def\mn@eprint@dblp#1{\href {http://dblp.uni-trier.de/rec/bibtex/#1.xml}
  {dblp:#1}}
\def\mn@eprint@#1:#2:#3:#4\@nil{\def\@tempa {#1}\def\@tempb {#2}\def\@tempc
  {#3}\ifx \@tempc \@empty \let \@tempc \@tempb \let \@tempb \@tempa \fi \ifx
  \@tempb \@empty \def\@tempb {arXiv}\fi \@ifundefined
  {mn@eprint@\@tempb}{\@tempb:\@tempc}{\expandafter \expandafter \csname
  mn@eprint@\@tempb\endcsname \expandafter{\@tempc}}}

\bibitem[\protect\citeauthoryear{Botteon et~al.,}{Botteon
  et~al.}{2018}]{botteon18a1758}
Botteon A.,  et~al., 2018, \mn@doi [MNRAS] {10.1093/mnras/sty1102}, 478, 885

\bibitem[\protect\citeauthoryear{Botteon et~al.,}{Botteon
  et~al.}{2019}]{botteon19lyra}
Botteon A.,  et~al., 2019, \mn@doi [A\&A] {10.1051/0004-6361/201936022}, 630,
  A77

\bibitem[\protect\citeauthoryear{Botteon et~al.,}{Botteon
  et~al.}{2020}]{botteon20a2255}
Botteon A.,  et~al., 2020, \mn@doi [ApJ] {10.3847/1538-4357/ab9a2f}, 897, 93

\bibitem[\protect\citeauthoryear{Brunetti \& Jones}{Brunetti \&
  Jones}{2014}]{brunetti14rev}
Brunetti G.,  Jones T.,  2014, \mn@doi [IJMPD] {10.1142/S0218271814300079}, 23,
  30007

\bibitem[\protect\citeauthoryear{Brunetti \& Vazza}{Brunetti \&
  Vazza}{2020}]{brunetti20}
Brunetti G.,  Vazza F.,  2020, \mn@doi [Phys. Rev. Lett.]
  {10.1103/PhysRevLett.124.051101}, 124, 51101

\bibitem[\protect\citeauthoryear{David \& Kempner}{David \&
  Kempner}{2004}]{david04}
David L.,  Kempner J.,  2004, \mn@doi [ApJ] {10.1086/423195}, 613, 831

\bibitem[\protect\citeauthoryear{Govoni et~al.,}{Govoni
  et~al.}{2019}]{govoni19}
Govoni F.,  et~al., 2019, \mn@doi [Science] {10.1126/science.aat7500}, 364, 981

\bibitem[\protect\citeauthoryear{Haines et~al.,}{Haines
  et~al.}{2018}]{haines18}
Haines C.,  et~al., 2018, \mn@doi [MNRAS] {10.1093/mnras/sty651}, 477, 4931

\bibitem[\protect\citeauthoryear{Intema, van~der Tol, Cotton, Cohen, van Bemmel
   \& R{\"{o}}ttgering}{Intema et~al.}{2009}]{intema09}
Intema H.,  van~der Tol S.,  Cotton W.,  Cohen A.,  van Bemmel I.,
  R{\"{o}}ttgering H.,  2009, \mn@doi [A\&A] {10.1051/0004-6361/200811094},
  501, 1185

\bibitem[\protect\citeauthoryear{Monteiro-Oliveira, Cypriano, Machado, {Lima
  Neto}, Ribeiro, Sodr{\'{e}}  \& Dupke}{Monteiro-Oliveira
  et~al.}{2017}]{monteirooliveira17a1758}
Monteiro-Oliveira R.,  Cypriano E.,  Machado R.,  {Lima Neto} G.,  Ribeiro A.,
  Sodr{\'{e}} L.,   Dupke R.,  2017, \mn@doi [MNRAS] {10.1093/mnras/stw3238},
  466, 2614

\bibitem[\protect\citeauthoryear{Murgia, Govoni, Feretti  \& Giovannini}{Murgia
  et~al.}{2010}]{murgia10}
Murgia M.,  Govoni F.,  Feretti L.,   Giovannini G.,  2010, \mn@doi [A\&A]
  {10.1051/0004-6361/200913414}, 509, A86

\bibitem[\protect\citeauthoryear{Offringa \& Smirnov}{Offringa \&
  Smirnov}{2017}]{offringa17}
Offringa A.,  Smirnov O.,  2017, \mn@doi [MNRAS] {10.1093/mnras/stx1547}, 471,
  301

\bibitem[\protect\citeauthoryear{Offringa et~al.,}{Offringa
  et~al.}{2014}]{offringa14}
Offringa A.,  et~al., 2014, \mn@doi [MNRAS] {10.1093/mnras/stu1368}, 444, 606

\bibitem[\protect\citeauthoryear{{Planck Collaboration XXVII}}{{Planck
  Collaboration XXVII}}{2016}]{planck16xxvii}
{Planck Collaboration XXVII} 2016, \mn@doi [A\&A]
  {10.1051/0004-6361/201525823}, 594, A27

\bibitem[\protect\citeauthoryear{Schellenberger, David, O'Sullivan, Vrtilek  \&
  Haines}{Schellenberger et~al.}{2019}]{schellenberger19}
Schellenberger G.,  David L.,  O'Sullivan E.,  Vrtilek J.,   Haines C.,  2019,
  \mn@doi [ApJ] {10.3847/1538-4357/ab35e4}, 882, 59

\bibitem[\protect\citeauthoryear{Shimwell et~al.,}{Shimwell
  et~al.}{2019}]{shimwell19}
Shimwell T.,  et~al., 2019, \mn@doi [A\&A] {10.1051/0004-6361/201833559}, 622,
  A1

\bibitem[\protect\citeauthoryear{Smirnov \& Tasse}{Smirnov \&
  Tasse}{2015}]{smirnov15}
Smirnov O.,  Tasse C.,  2015, \mn@doi [MNRAS] {10.1093/mnras/stv418}, 449, 2668

\bibitem[\protect\citeauthoryear{Tasse}{Tasse}{2014}]{tasse14arx}
Tasse C.,  2014, arXiv e-prints

\bibitem[\protect\citeauthoryear{Tasse et~al.,}{Tasse et~al.}{2018}]{tasse18}
Tasse C.,  et~al., 2018, \mn@doi [A\&A] {10.1051/0004-6361/201731474}, 611, A87

\bibitem[\protect\citeauthoryear{Vazza, Ettori, Roncarelli, Angelinelli,
  Br{\"{u}}ggen  \& Gheller}{Vazza et~al.}{2019}]{vazza19}
Vazza F.,  Ettori S.,  Roncarelli M.,  Angelinelli M.,  Br{\"{u}}ggen M.,
  Gheller C.,  2019, \mn@doi [A\&A] {10.1051/0004-6361/201935439}, 627, A5

\bibitem[\protect\citeauthoryear{de Gasperin et~al.,}{de~Gasperin
  et~al.}{2019}]{degasperin19}
de Gasperin F.,  et~al., 2019, \mn@doi [A\&A] {10.1051/0004-6361/201833867},
  622, A5

\bibitem[\protect\citeauthoryear{van Haarlem et~al.,}{van Haarlem
  et~al.}{2013}]{vanhaarlem13}
van Haarlem M.,  et~al., 2013, \mn@doi [A\&A] {10.1051/0004-6361/201220873},
  556, A2

\bibitem[\protect\citeauthoryear{van Weeren, de Gasperin, Akamatsu,
  Br{\"{u}}ggen, Feretti, Kang, Stroe  \& Zandanel}{van Weeren
  et~al.}{2019}]{vanweeren19rev}
van Weeren R.,  de Gasperin F.,  Akamatsu H.,  Br{\"{u}}ggen M.,  Feretti L.,
  Kang H.,  Stroe A.,   Zandanel F.,  2019, \mn@doi [Space Sci. Rev.]
  {10.1007/s11214-019-0584-z}, 215, 16

\makeatother
\end{thebibliography}

\bsp	
\label{lastpage}
\end{document}